\documentclass[aps,showpacs,twocolumn,amsmath]{revtex4-1}
\usepackage{amsmath}
\usepackage{txfonts}
\usepackage{hyperref}
\usepackage{graphicx}
\usepackage{float}
\usepackage{xcolor}
\usepackage[normalem]{ulem}
\usepackage{todonotes}

\newcommand{\Var}{\mathrm{Var}}
\newcommand{\lk}[1]{\textcolor{black}{#1}}

\begin{document}

\title{On subdiffusive continuous time random walks with stochastic resetting}
\author{\L{}ukasz Ku\'smierz}
\affiliation{Laboratory for Neural Computation and Adaptation, RIKEN Center for Brain Science, 2-1 Hirosawa, Wako, Saitama 351-0198, Japan}
\author{Ewa Gudowska-Nowak}
\affiliation{Marian Smoluchowski Institute of Physics, Jagiellonian University, ul. \L{}ojasiewicza 11, 30-348 Krak\'ow, Poland}
\affiliation{Mark Kac Complex Systems Research Center, Jagiellonian University, Krak\'ow, Poland}
\begin{abstract}
We analyze two models of subdiffusion with stochastic resetting. 
Each of them consists of two parts: 
subdiffusion based on the continuous-time random walk (CTRW) 
scheme and independent resetting events generated uniformly 
in time according to the Poisson point process.
In the first model the whole process is reset to the initial state, 
whereas in the second model only the position is subject to resets.
The distinction between these two models arises from 
the non-Markovian character of the subdiffusive process.  
We derive exact expressions for the two lowest moments 
of the full propagator, stationary distributions, 
and first hitting times statistics. 
We also show, with an example of a constant drift,
how these models can be generalized to include external forces. 
Possible applications to data analysis and 
modeling of biological systems are also discussed.
\end{abstract}

\maketitle
\section{Introduction}
\label{sec:introduction}
Recently a simple model of diffusion with stochastic resetting has been 
proposed \cite{majumdar2011resetting}. 
In its basic form it is a Brownian motion interrupted by reset events, 
which instantaneously bring the process back to the initial position. 
Resets happen randomly in time, according to the Poisson point 
process with intensity $r$. 
In contrast to standard, symmetric diffusion, 
diffusion with resetting leads to a finite 
mean first arrival time (MFAT) to any fixed position 
\cite{majumdar2011resetting}. 
Moreover, it has been shown that resetting may be beneficial for search in 
a variety of modified scenarios, e.g. 
when the initial random walk process is superdiffusive 
\cite{kusmierz2014first,kusmierz2015optimal}, 
when the space is higher-dimensional 
\cite{evans2014diffusion} and/or bounded \cite{christou2015diffusion},
and when the resetting intensity is inhomogeneous in time 
\cite{Reuveni2016,pal2016first,eule2016non,bhat2016stochastic,pal2016diffusion,kusmierz2018robust,maso2019stochastic,maso2019transport}. 
The combination of a random walk process with 
stochastic resetting events leads to many nontrivial features: 
A non-local character of resets results in a current-carrying 
non-equilibrium steady states (NESS), 
i.e. stationary distributions in which detailed balance does not hold
\cite{majumdar2011resetting,evans2013optimal,pal2015diffusion,eule2016non,montero}.
When analyzed as a search process involving a mixture of local steps with resets, 
the mean first passage time to find a target has been computed exactly and
has been found to have a minimum at an optimal resetting rate \cite{majumdar2011resetting}, 
indicating that the search process is more efficient in the presence of resetting. 
This non-equilibrium search is more efficient than a corresponding 
(leading to the same shape of the stationary distribution) conservative-force-induced equilibrium dynamics \cite{evans2013optimal} and 
in some cases more efficient than any conservative-force-induced equilibrium dynamics \cite{kusmierz2017optimal}.
When instead of the Brownian motion the search is performed 
by long range L\'evy moves combined with resets, 
the motion generates a bifurcation in the optimal search strategy 
(i.e. bifurcation in the shortest mean first-time to reach a target). 
The bifurcation is discontinuous \cite{kusmierz2015optimal} or continuous \cite{kusmierz2014first}, 
depending on whether the process is performed in discrete or continuous time. 
Stochastic resets can modify the splitting probabilities 
leading to a higher success rate in Bernoulli trials \cite{belan2018restart}. 
A one-dimensional version of the walk where the walker, at each time step, 
resets to the maximum of the already visited positions with a certain probability leads to 
a dynamical transition in the temporal relaxation \cite{majumdar2015dynamical}.
The stochastic resetting model is related to reinforced random walks \cite{davis1990reinforced},   
a version of which has recently been proposed as a model of large-scale movement of free-ranging capuchin monkey 
and shown to lead to super-slow (logarithmic) diffusion \cite{boyer2014random}.
It can also lead to a localization transition similar to Anderson localization 
if supplemented by a simple model of non-homogeneous learning \cite{falcon2017localization}.

Here we explore another generalization of diffusion with resetting, 
wherein the diffusive process between resets is substituted with a process 
generated by the continuous-time random walk (CTRW) scheme 
\cite{montroll1965random,barkai2000continuous,meerschaert2004limit,metzler2000random}. 
In contrast to recent works that combine stochastic resets with the CTRW with 
exponentially distributed waiting times between jumps
\cite{montero,campos2015phase,Montero2017}, 
we assume that the waiting times are distributed according to 
a power law $p(t)\propto t^{-\mu-1}$, where $t\gg1$ and $\mu\in(0,1)$.
Due to the infinite mean waiting time the primary process (i.e. without resetting) 
is subdiffusive with the mean square displacement (MSD) 
growing sublinearly with time, 
$\langle X^2(t)\rangle \propto t^{\mu}$, 
and the Brownian diffusion is recovered in the limit $\mu \to 1$.
A subdiffusive CTRW process is non-Markovian \cite{scalas2000fractional} 
and shows weak egodicity breaking \cite{bel2005weak,lubelski2008nonergodicity}, 
i.e. time and ensemble averages of physical observables do not coincide. 

In addition to the CTRW process there are several distinct
mathematical models of subdiffusion \cite{sokolov2015}, 
differing significantly in the details of dynamics.
Two most popular are fractional Brownian motion (FBM)
\cite{mandelbrot1968fractional,taqqu1975weak,biagini2008stochastic,mishura2008stochastic,deng2009ergodic}
and diffusion on fractals
\cite{d1983random,o1985diffusion,ben2000diffusion}.
Each of these models corresponds to a different physical mechanism
leading to the subdiffusive behavior.
While all of them share the sublinear growth of the variance,
they exhibit many dissimilarities in other features.
Importantly, in contrast to the subdiffusive CTRW, 
the other two processes are ergodic (but see \cite{deng2009ergodic}).  
Recently, statistical methods of discriminating between FBM and CTRW
\cite{magdziarz2009fractional}, diffusion on fractals and CTRW
\cite{condamin2008probing}, and diffusion on fractals and FBM
\cite{meroz2013test} have been proposed.

Subdiffusion has been observed in a number of distinct physical phenomena 
\cite{scher1975a,kopelman1984reaction,scher2002dynamical,weiss2007anomalous}. 
Many experiments have shown that the transport in living cells is subdiffusive
\cite{schwille1999molecular,wachsmuth2000anomalous,tolic2004anomalous}, 
but the underlying mechanism is not yet fully understood. 
In this context, some recent works report non-ergodic CTRW-like behavior 
\cite{jeon2011vivo,weigel2011ergodic,jeon2012anomalous,tabei2013intracellular}, 
whereas others are in favor of ergodic FBM or diffusion on fractals
\cite{szymanski2009elucidating,weber2010bacterial,kepten2011ergodicity,barkai2012single}. 
In view of those findings, we study how the statistical properties of 
a subdiffusive CTRW process are altered under random relocations. 
The overall process is a combination of a CTRW and independent resets. 
In subsequent Sections we define two models of the resetting 
mechanism when either relocation event erases memory about 
the past motion (Model I) or, otherwise resets do not interfere
with waiting for the next step of the random motion (Model II).
As we will show, non-trivial features of such processes may not only 
provide a descriptive model of natural phenomena, 
such as dynamics of molecular motors in the crowded cytoplasm, 
but also may be helpful in designing resampling-based 
statistical inference methods aiming at 
detecting trapping and associated ergodicity breaking.

Some of our results can be considered to be special cases of previous more general 
considerations \cite{pal2016first,shkilev2017continuous,maso2019transport}. 
In contrast to these works, 
 we focus almost exclusively on the CTRW subdiffusion with exponential resetting 
and analyze the possible outcomes in depth, uncovering intriguing effects 
that, to best of our knowledge, have not yet been reported. 
Power-law waiting times have also been recently explored in \cite{evans2018effects,maso2019stochastic} 
in the context of refractory periods/residence times 
that follow resets. 
Such trapping at the resetting position is related to, 
but distinct from the space-homogeneous trapping which happens in our Model I. 
Interestingly, our Model II also has a net effect 
of trapping at the resetting position. 
Notice, however, that the underlying mechanism is different from models in 
\cite{evans2018effects,maso2019stochastic} and is driven by the power-law 
distributed waiting times between steps.

Throughout the paper we use the following notation. 
The propagator of a process of interest is written as 
${W(x,t|x_0)\equiv W(x,t|x_0,0)}$, with the initial condition
$W{(x,0|x_0) = \delta(x-x_0)}$. 
Partial derivatives with respect to time and space are written as 
$\partial_t$ and $\partial_x$, respectively.
The parameter $r$, called resetting intensity, gives the expected number of 
reset events in unit time.
We denote the space Fourier transform and the time Laplace transform implicitly, 
by changing the argument(s) of the transformed function 
($x\to k$ and $t\to s$, respectively).

\section{Subdiffusion and Fractional Fokker-Planck equation}
Here we list basic results related to CTRW subdiffusion that we will use throughout the paper. 
For a complete description, see e.g. \cite{metzler2000random}. 
In our derivations of quantities related to the process with resetting we 
will often employ the function  
\begin{equation}
W_0(x,s|x_0) = 
\frac{s^{\frac{\mu}{2}-1}}{2 \sqrt{D}}e^{-\sqrt{\frac{s^{\mu}}{D}}|x-x_0|}, 
\label{eq:propagatorW0}
\end{equation}
which is the (Laplace transformed) solution of a time-fractional Fokker-Planck (FFPE) equation
\begin{equation}
\partial_t W_0(x,t|x_0) 
=
{}_0 D{_t^{1-\mu}} D \partial_{xx}^2 W_0(x,t|x_0),
\label{eq:W0equation}
\end{equation}
where ${}_0 D{_t^{1-\mu}}$ denotes the Riemann-Liouville fractional derivative 
operator \cite{metzler1999anomalous,metzler2000random,metzler2000subdiffusive} {with $0<\mu<1$}.
Function $W_0(x,t|x_0)$ is the propagator of the process without resetting, 
i.e. probability density function (PDF) of the CTRW subdiffusion 
starting from $x_0$ at time $t_0=0$.
Multiplying (\ref{eq:propagatorW0}) by $(x-x_0)^2$ and integrating over $x$ 
we obtain the variance (in this case equivalent to the MSD) 
evolution in the Laplace space 
$\langle X^2(s)\rangle = 2 D/s^{1+\mu}$ which translates to
\begin{equation}
    \langle X^2(t)\rangle = \frac{2 D}{\Gamma(\mu + 1)} t^{\mu}.
\end{equation}
The FFPE (\ref{eq:W0equation}) can be derived as follows. 
The movement of a particle is generated via jumps generated from a PDF $g(x)$. 
The time between two consecutive jumps is generated from a non-negative random variable, called waiting times, 
that is described by a PDF $\psi(t)$ or, equivalently, by the {probability that the time before the next jump happened is longer than $t$, }
$\Psi(t) = \int_t^{\infty} d \tau \psi(\tau)$.   
Between subsequent jumps the particle is immobile and, importantly, jumps and waiting times are independent. 
It is well known that the propagator of such process is given by the Montroll-Weiss formula 
\cite{montroll1965random,metzler2000random,zaburdaev2015levy}
\begin{equation}
    W_0(k,s|x_0) = \frac{\Psi(s) e^{i k x_0}}{1 - \psi(s)g(k)}.
    \label{eq:MontrollWeiss}
\end{equation}
In the case of the subdiffusive CTRW, we assume that ${g(k)\approx 1 - D_0 k^2 }$ 
and ${\psi(s) \approx 1 - A_{\mu} s^{\mu}}$ 
 and perform the limit 
${A_{\mu}\to 0}$, ${D_0 \to 0}$ with ${D_0/A_{\mu} \to D}$, which leads to 
(\ref{eq:propagatorW0}) and (\ref{eq:W0equation}).
\lk{Most results obtained in this paper will have two forms: 
a general CTRW form generalizing (\ref{eq:MontrollWeiss}) 
and a subdiffusive form that follows from taking the described limit. 
In some cases this will allow us to write a modified 
fractional Fokker-Planck equation. 
Nevertheless, our derivations are based on the renewal approach that 
can be used in the very general setting of the CTRW.
}

It is easy to calculate the Laplace transform of the survival 
probability $S_0(t)$ of the subdiffusive search process without resetting {
in the presence of a single target located at $x=0$}. 
The corresponding unnormalized density $G_0(x,t|x_0)$ is the solution of the equation
\begin{equation}
\partial_t G_0(x,t|x_0) 
=
{}_0 D{_t^{1-\mu}} D \partial_{xx}^2 G_0(x,t|x_0),
\label{eq:G0equation}
\end{equation}
which has the same form as (\ref{eq:W0equation}) except for the boundary conditions. 
The target is introduced by imposing the {absorption condition} ${G_0(x=0,t|x_0) = 0}$.
We can construct such a solution from the solution (\ref{eq:propagatorW0}) of 
(\ref{eq:W0equation}) via the method of images, leading to
\begin{equation}
    G_0(x,s|x_0)) = \frac{s^{\mu/2-1}}{2\sqrt{D}}
    \left(e^{-\sqrt{s^{\mu}/D}|x-x_0|} -
    e^{-\sqrt{s^{\mu}/D}|x+x_0|} \right).
    \label{eq:G0solution}
\end{equation}
The corresponding survival probability reads 
\begin{equation}
    S_0(s) = \int\limits_0^{\infty} \mathrm{d}x G_0(x,s|x_0) = 
    \frac{1}{s}\left(
    1 - e^{- \sqrt{s^{\mu}/D}|x_0|}
    \right), 
    \label{eq:S0}
\end{equation}
which shows that, despite the trapping, in the lomg time limit 
$\lim_{t\to\infty} S_0(t) = 0$, i.e. 
the target is almost surely (with probability $1$) reached. 
The {first arrival times (FAT)} probability density function 
\begin{equation}
    \rho_0(s) = 1 - s S_0(s) = \exp\left(-\sqrt{s^{\mu}/D}|x_0|\right)
    \label{eq:rho0}
\end{equation}
behaves like $\propto t^{-(1+\mu/2)}$ for large $t$, 
which generalizes the well-known Sparre Andersen scaling $\propto t^{-3/2}$, 
which holds for symmetric Markovian processes with stationary increments
\cite{andersen1954fluctuations,andersen1955fluctuations,dybiec2009anomalous}. 

\section{First model: resetting of the process}
Resets in the first model bring the position to the origin and delete 
all memory about the past. In the CTRW scheme waiting times for the diffusive 
jump and for the reset are drawn independently. 
Whenever the latter is shorter, the position is reset to $0$ 
and both waiting times are generated again. 
Thus, the aging effect of subdiffusion is also subject to reset.
In the following we will refer to such events as hard resets. 
Although we assume that resetting brings the process back to the initial position, 
it is convenient to perform calculations for any initial position $x_0$, 
with the resetting position kept at the origin. 
At the end we set $x_0 = 0$.

Given these assumptions the propagator of the process subject to 
random resets can be written in terms of the propagator of 
the process without resetting via renewal equation
(\cite{gupta2014fluctuating,kusmierz2015optimal}) 
\begin{equation}
    W(x,t|x_0) = 
    e^{-r t} W_0(x,t|x_0) + 
    \int\limits_0^{{t}}d\tau r e^{-r \tau} W_0(x,\tau|{0}),
\end{equation}
which in the Laplace space reads
\begin{equation}
W(x,s|x_0)=W_0(x,r+s|x_0)+\frac{r}{s}W_0(x,r+s|0).
\label{eq:WfromW0}
\end{equation}
We can therefore use the known propagator of the CTRW process (\ref{eq:MontrollWeiss}) to 
construct the propagator of the process with hard resets, 
which reads
\begin{equation}
W(k,s|x_0) = 
\frac{\Psi(r+s)\left(e^{i k x_0} + r/s\right)}{1 - \psi(r+s)g(k)}.
\label{eq:propagator-hard-resets}
\end{equation}
\lk{
Expression (\ref{eq:propagator-hard-resets}) is very general and 
can be used to study any CTRW process with hard resets. 
In fact, it would be straightforward to generalize it further to include spatio-temporal coupling between the jump distribution and waiting times, 
as in the case of L\'evy walks \cite{zaburdaev2015levy}. 
As announced, here we focus on the special case of subdiffusive CTRW.  
}

\lk{
We note in passing that, for any $g(x)$ that is continuous at $x=0$, 
the propagator (\ref{eq:propagator-hard-resets}) 
has a singular part that behaves like $\Psi(r+s)(e^{i k x_0} + r/s)$. 
However, if we assume that ${g(k)\approx 1 - D_0 k^2 }$ 
and ${\psi(s) \approx 1 - A_{\mu} s^{\mu}}$ 
with 
${A_{\mu}\to 0}$, ${D_0 \to 0}$, and ${D_0/A_{\mu} \to D}$, 
the singular part disappears. 
As we will show later, this is one of the features that discriminates 
hard from soft (Model II) resets. 
}

\lk{
Taking such limit is equivalent to plugging the subdiffusive 
$W_0$ given by (\ref{eq:propagatorW0}) into (\ref{eq:WfromW0}). 
}
The corresponding propagator with hard resets in the Fourier-Laplace space $W(k,s|x_0)$ solves the following algebraic equation
\begin{equation}
\left( D \mathbin{{(s+r)^{1-\mu}}}k^2+s+r\right) W(k,s|x_0) = 
e^{i k x_0}+\frac{r}{s}.
\label{Eq:FFPE1}
\end{equation}
Eq. (\ref{Eq:FFPE1}) corresponds to the following integro-differential 
equation in $(x,t)$ space
\begin{equation}
\partial_t W(x,t|x_0) = 
D e^{-r t} {}_0 D{_t^{1-\mu}} e^{r t}\partial_{xx}^2 W(x,t|x_0) - r W(x,t|x_0) + r\delta(x).
\end{equation} 
The operator $e^{-r t} {}_0 D{_t^{1-\mu}} e^{r t}$ 
corresponds to a version of truncated Riemann-Liouville fractional operator 
\cite{meerschaert2008tempered,gajda2010fractional,alrawashdeh2017applications}.
The same operator was previously proposed in the context of reaction-subdiffusion 
\cite{schmidt2007mesoscopic} and, 
interestingly, it also appears in the context of 
subdiffusion with a constant external force that affects 
waiting times \cite{fedotov2015subdiffusion}.
In our case, it emerges because the resets induce two effects: 
the position of the particle is set to $0$ 
and the waiting times are generated anew. 
The latter phenomenon introduces an effective truncation on the waiting times, 
which is realized in the Fractional Fokker-Planck equation through 
the described modification of the Riemann-Liouville fractional operator. 
From Eq. (\ref{Eq:FFPE1}) we obtain the stationary distribution
\begin{equation}
f_s(x) = 
\lim_{t\to \infty}W(x,t|x_0)=
\lim_{s\to 0} s W(x,s|x_0) =
\frac{1}{2}\sqrt{\frac{r^{\mu}}{D}} e^{-\sqrt{\frac{r^{\mu}}{D}}|x|}.
\label{Eq:stationary1}
\end{equation}
{Recently the same formula has been independently derived 
in \cite{maso2019transport}.}
As expected, for $\mu = 1$ we recover formula derived for diffusion with 
stochastic resetting \cite{majumdar2011resetting}. 
For any $\mu$ we obtain the Laplace distribution. 
The difference lies in the dependence of its mean value on the resetting 
intensity, which could also be deduced from dimensional analysis. 

In order to examine the transient behavior of the process 
we calculate first and second moments. 
First, we focus on the general case of a process 
starting from some position $x_0$. 
By inspecting Eq. (\ref{Eq:FFPE1}) it is easy to check that 
$\langle X(t) \rangle = x_0 e^{-r t}$, 
i.e. relaxation of the expected position is exponential 
and is driven purely by the resetting events.
Relaxation of the variance incorporates two distinct processes. 
The first of them is related to relaxation of the position, 
and is described by the expression $(e^{-r t}-e^{-2 r t})x_0^2$.
As announced before, we focus on the second relaxation process, 
which is strictly related to the width change, and thus we set $x_0=0$, 
i.e. we assume that the process starts from the resetting position. 
From Eq. (\ref{Eq:FFPE1}) we obtain the MSD in the Laplace space
\begin{equation}
\langle X^2(s) \rangle = 
\frac{2 D}{s (s+r)^{\mu}}.
\end{equation}
Inverting the Laplace transform leads to the integral expression
\begin{equation}
\mbox{Var} X(t) = 
\langle X^2(t) \rangle =  
\frac{2 D}{\Gamma(\mu)}\int\limits_0^t d\tau \tau^{\mu-1}e^{-r \tau} = 
\frac{2 D}{ r^{\mu}\Gamma(\mu)}\gamma(\mu,r t),
\end{equation}
where $\gamma(s,x)$ stands for the lower incomplete gamma function. 
Asymptotically 
\begin{equation}
    \langle X^2(t)\rangle \approx 
    \frac{2 D}{r^{\mu}}\left( 1 - \frac{e^{-rt}}{\Gamma(\mu)(rt)^{1-\mu}}\right),
\end{equation} 
as $t\to\infty$, 
i.e. the stationary state is approached exponentially fast. 
\begin{figure}
\includegraphics[width=1.0\linewidth]{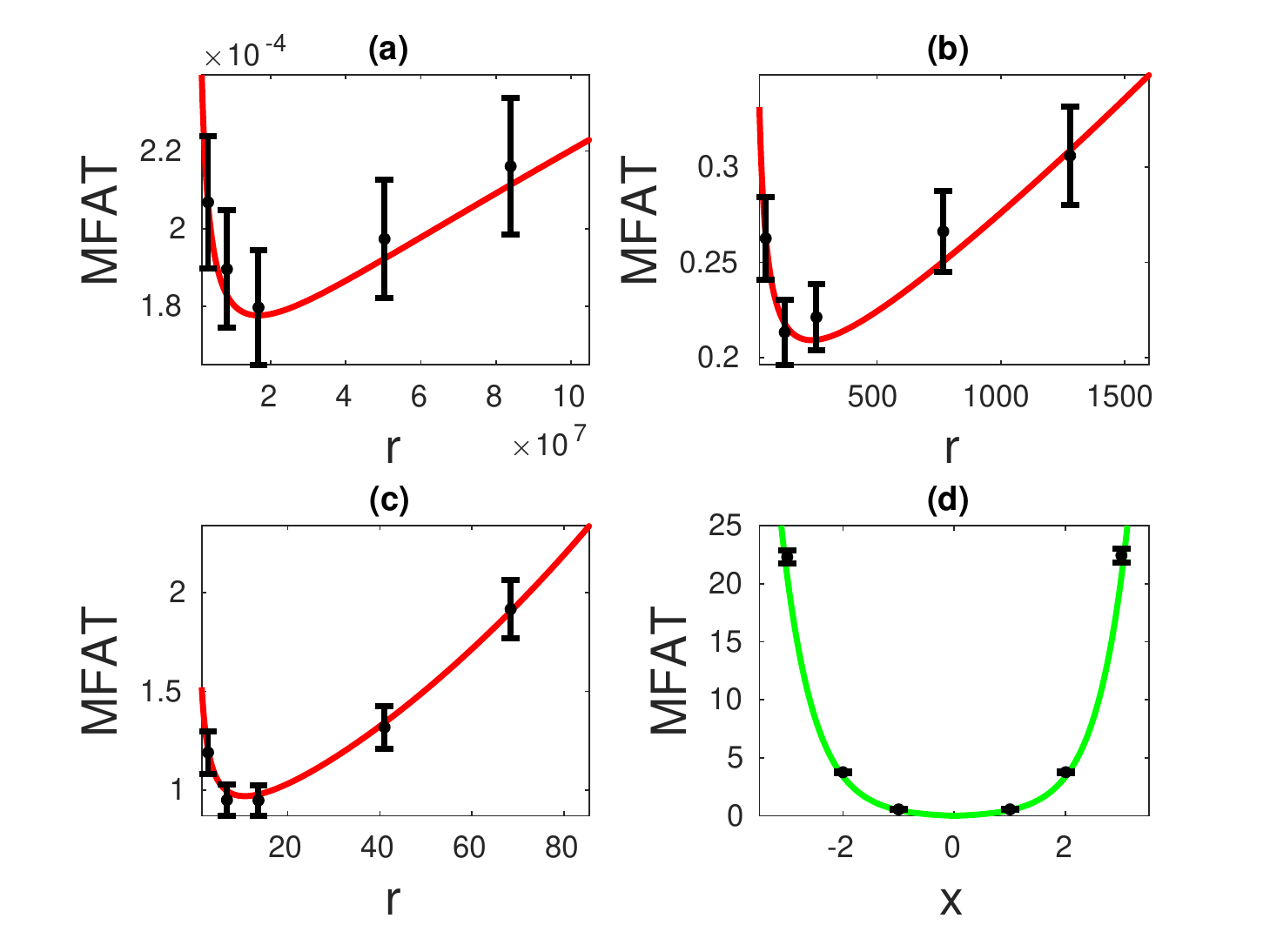}
\caption{Comparison of the mean first arrival time 
(MFAT$\equiv\langle T_r\rangle$) 
obtained with analytical prediction 
(Eq. (\ref{Eq:MFATFiniteSub}), lines) and stochastic simulations 
(points with $99\%$ confidence intervals) 
as a function of resetting intensity (\lk{from (a) to (c),} 
red, $\mu=\{0.25, 0.5, 0.75\}$) 
and target position (\lk{(d),} green, $\mu=0.5$). }
\label{Fig:model1_numerics}
\end{figure}

In the following we calculate the first arrival time (FAT) statistics. 
We follow the general strategy introduced in \cite{pal2016first}. 
Due to the renewal structure of the resetting, 
we can link the FAT distribution of the process with resetting with 
$S_0$. 
We introduce a random variables $R$ (time to the next reset), 
$T_0$ (FAT of the process without resetting), 
and $T_r$ (FAT of the process with resetting).
The resets can be introduced as follows: 
first draw $T_0$ and $R$. 
If $T_0\leq R$ then the reset was supposed to happen after the target was found, 
thus $T_r = T_0$. If, on the other hand, $T_0>R$, then the reset came before the target was 
supposed to be found, thus the search has to start anew. 
In that case $T_r = R + T_r'$, 
where $T_r'$ is a new realization of the FAT of the process with resetting, 
which in turn depends on the independent realizations of 
the resetting times and FATs of the process without resetting. 
We can summarize this procedure with the equation
\begin{equation}
    \begin{split}
        T_r &= I[T_0\leq R] T_0 + I[T_0 > R] (R + T_r') 
 \\    &= \min(T_0,R) + I[T_0 > R]T_r',
    \end{split}
    \label{eq:renewalT}
\end{equation}
where the indicator function ${I[a>b] = 1}$ if ${a>b}$ and ${I[a>b]=0}$ if ${a\leq b}$.
Since $T_r'$ is independent from $T_0$ and $R$, we can easily average both sides of 
(\ref{eq:renewalT}) leading to the known general formula for the MFAT 
of the process subject to resetting \cite{pal2016first}
\begin{equation}
    \langle T_r \rangle = \frac{\langle \min(T_0,R)\rangle}{\langle I[T_0\leq R]\rangle},
    \label{eq:generalMeanT}
\end{equation}
where averaging is performed over independent $R$ and $T_0$. 
In the case of exponential (i.e. constant rate) resetting 
${\langle \min(R,T_0) \rangle = S_0(r)}$ 
and ${\langle I[T_0 \leq R] \rangle = 1 - r S_0(r)}$, thus 
\begin{equation}
    \langle T_r \rangle = \frac{S_0(r)}{1 - r S_0(r)},
    \label{eq:exponentialMeanT}
\end{equation}
where $S_0(r)\equiv S_0(s=r)$. 
Combining (\ref{eq:S0}) and (\ref{eq:exponentialMeanT}) we arrive at 
\begin{equation}
\langle T_{r}\rangle = 
\frac{1}{r} \left(e^{\sqrt{{r^{\mu}}/{D}}|x_0|}-1\right).
\label{Eq:MFATFiniteSub}
\end{equation}
Equation (\ref{Eq:MFATFiniteSub}) is consistent with a more general formula 
derived with different methods in \cite{shkilev2017continuous}. 
In the limit of $\mu\to1$ it simplifies to the well-known formula for the MFAT 
of diffusion with stochastic resetting \cite{majumdar2011resetting}.
In contrast to superdiffusive L\'evy flights with resetting 
\cite{kusmierz2014first,kusmierz2015optimal}, 
the CTRW subdiffusion with resetting preserves the exponential form of 
of the dependence of the MFAT on the distance to the target. 
Interestingly, for a fixed $|x_0|>0$, 
there is no local minimum in parameter space $(r,\mu)$. 
The infimum of the MFAT is $0$, 
\begin{equation}
\lim_{r=\frac{1}{\mu}\to \infty}\left\langle T_{r} \right\rangle=0,
\end{equation}
because subdiffusion is slow at large time-scales, 
but extremely fast at short time-scales. 
This can be seen from the following formula, 
which holds for pure subdiffusion without resetting ($r=0$)
\begin{equation}
\partial_t \langle X^2(t) \rangle \propto t^{\mu-1}.
\end{equation}
Note, however, that at short time-scales this formula cannot describe physical 
phenomena for which the speed is bounded. 

For a fixed $\mu$ there is exactly one minimum of the MFAT as a function of $r$.
The optimal resetting intensity is given by the expression
\begin{equation}
r^*=\left(\frac{z^2_{\mu} D}{x^2}\right)^{\frac{1}{\mu}},
\end{equation}
where $z_{\mu}$ is the solution of the transcendental equation
\begin{equation}
\frac{\mu}{2} z_{\mu} = 1-e^{-z_{\mu}}.
\label{eq:transc}
\end{equation}
When $z_{\mu}$ is large, 
the exponential element on the RHS of Eq. (\ref{eq:transc}) is negligible, 
thus we can write an approximate solution
\begin{equation}
z_{\mu}\approx \frac{2}{\mu},
\end{equation}
which is valid for $\mu\ll1$. 
This leads to the following dependence of the optimal resetting intensity on $\mu$
\begin{equation}
r^* \propto \left(\frac{2}{\mu}\right)^{\frac{2}{\mu}}.
\label{Eq:limitr}
\end{equation}

\begin{figure}
\includegraphics[width=1.0\linewidth]{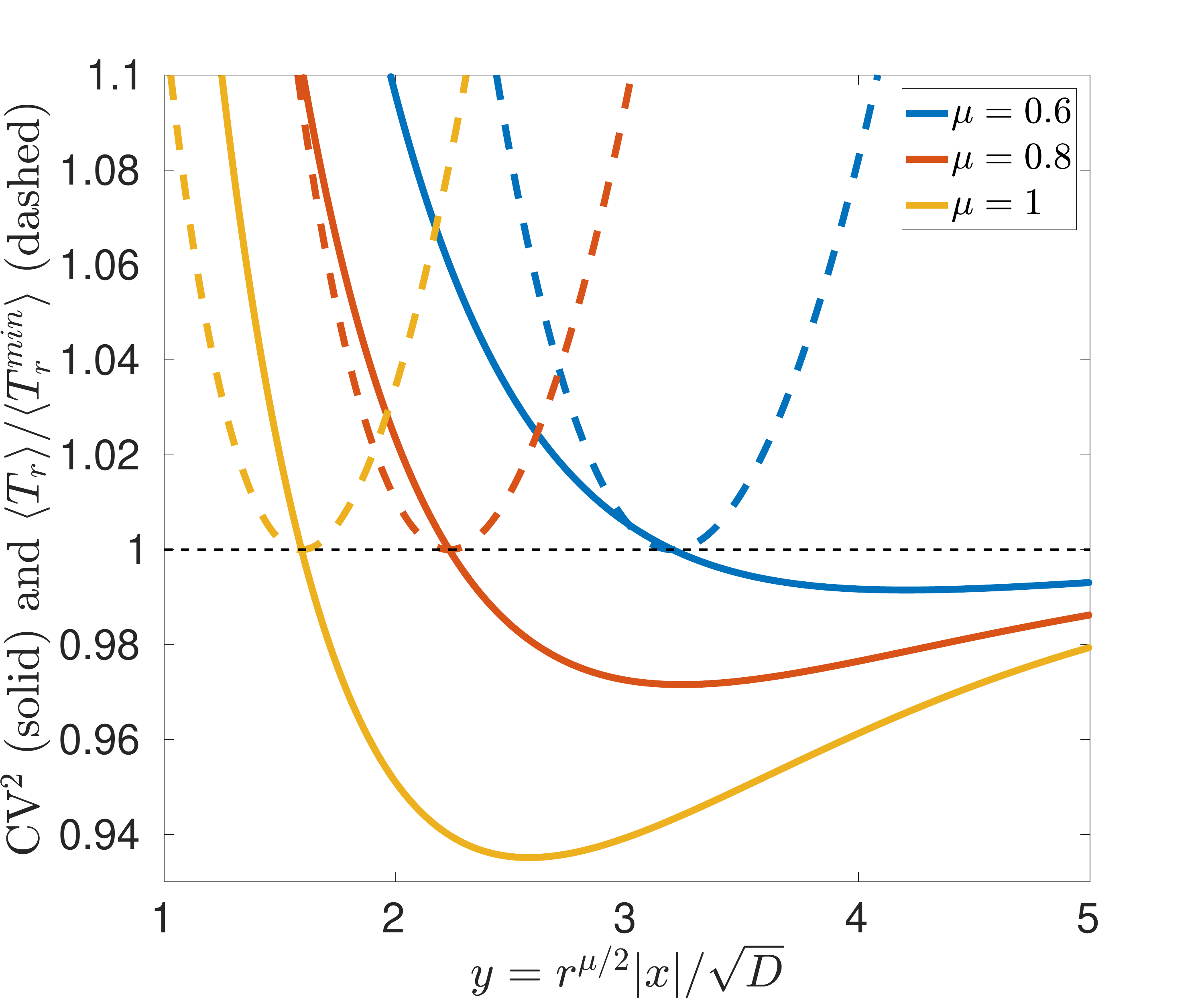}
\caption{The squared coefficients of variation (solid lines, (\ref{eq:cv2})) 
and the (scaled) mean first arrival times 
dashed lines, (\ref{Eq:MFATFiniteSub})) of subdiffusion with hard resets 
as functions of ${y=r^{\mu/2}|x|/\sqrt{D}}$ for different values of $\mu$ \lk{(increasing from left to right)}.
}
\label{Fig:cv}
\end{figure}

How is $T_{r}$ distributed around its mean value? 
We square both sides of (\ref{eq:renewalT}) and calculate averages 
\begin{equation}
    \langle \min(T_0,R)^2 \rangle = -2\partial_r S_0(r)
\end{equation}
and 
\begin{equation}
    \langle I[T_0 > R] \min(T_0,R) \rangle = - r \partial_r S_0(r),
\end{equation}
arriving at
\begin{equation}
    \Var T_r = \langle T_r^2 \rangle - \langle T_r\rangle^2 = 
    - \frac{2\partial_r S_0(r) + S_0(r)^2}{1 - r S_0(r)}.
    \label{eq:VarianceGeneral}
\end{equation}
Combining (\ref{eq:S0}) and (\ref{eq:VarianceGeneral}) we obtain
\begin{equation}
    \Var T_r = r^{-2} \left(
    e^{2 y} - \mu y e^y - 1
    \right),
    \label{eq:VarianceSubdiffusion}
\end{equation}
where $y=r^{\mu/2}|x_0|/\sqrt{D}$. 
Let us analyze the squared coefficient of variation.  
\begin{equation}
\mbox{CV}^2 = 
\frac{\mbox{Var} T_{r}}{\langle T_{r} \rangle^2} = 
1+\frac{2}{\left(e^{y}-1\right)^2}\left( \left(1-\frac{\mu}{2}y\right) e^{y}-1\right),
\label{eq:cv2}
\end{equation}
At $r=r^*$ (and thus $y=z_{\mu}$), i.e. when the intensity of resetting is optimal, 
the coefficient of variation is equal to $1$ (see Fig.~\ref{Fig:cv}), 
in line with the universal result obtained recently \cite{Reuveni2016}.
The resetting intensity at which $\mbox{CV}$ admits its minimum is given by
\begin{equation}
r^{**}=\left(\frac{y^2_{\mu} D}{x^2}\right)^{\frac{1}{\mu}},
\end{equation}
where $y_{\mu}$ is the solution of the transcendental equation
\begin{equation}
\frac{y_{\mu}}{1+\frac{2}{\mu}}=\tanh{\frac{y_{\mu}}{2}}.
\label{Eq:trans2}
\end{equation}
Note that $r^{**}>r^*$ for any $\mu\in(0,1]$ (cf. Fig.~\ref{Fig:cv}). 
For small $\mu$ we can approximate the solution of the transcendental equation 
(\ref{Eq:trans2}) by assuming $\tanh{\frac{y_{\mu}}{2}}\approx1$, leading to
\begin{equation}
y_{\mu}\approx1+\frac{2}{\mu}
\end{equation}
and thus
\begin{equation}
r^{**}\propto \left(1+\frac{2}{\mu}\right)^{\frac{2}{\mu}}.
\label{Eq:limitr2}
\end{equation}
Combining Eqs.~(\ref{Eq:limitr}) and (\ref{Eq:limitr2}) we can calculate the limit
\begin{equation}
\lim_{\mu\to0}\frac{r^{**}}{r^*}=e,
\end{equation}
which gives the upper bound on the relative separation between $r^*$ and $r^{**}$. 
We now show an alternative derivation of the FPT statistics, 
which should shed some light on the features of the 
CTRW subdiffusion with stochastic resetting. 
For a process starting at $t_0=0$ from $x_0=0$ we can link the distribution of the 
FAT to a position of the target $x$, denoted in the following as $\rho$, 
with the free propagator $W$ (without absorbing boundaries) \cite{montroll1973random,shlesinger1974asymptotic,scher1975anomalous,chechkin2003first,kusmierz2015optimal}
\begin{equation}
    W(x,t|x_0) = \int\limits_0^t \mathrm{d}\tau \rho(\tau) W(x,t-\tau|x).
    \label{eq:WandP}
\end{equation}
In the time-Laplace space (\ref{eq:WandP}) simplifies to an algebraic equation
\begin{equation}
    W(x,s|0) = \rho(s) W(x,s|x).
    \label{eq:WandPalg}
\end{equation}
We can now combine (\ref{eq:WfromW0}) and (\ref{eq:WandPalg}) to obtain the following 
formula for the Laplace transform of the FAT's probability density function in the process with hard resets
\begin{equation}
    \rho(s) = 
    \frac{s+r}{s (\rho_0(r + s))^{-1} + r}
    ,
    \label{eq:pfa1}
\end{equation}
with $\rho_0(s)$ standing for the (Laplace-transformed) PDF of first arrival times in the (corresponding) process free of resets. 
Equation (\ref{eq:pfa1}) has previously been derived \cite{Reuveni2016,pal2016first} 
using the same general technique that we used here to derive formulas for 
$\langle T_r \rangle$ and $\Var T_r$ and one can easily verify 
that it leads to the same expressions 
(plug in (\ref{eq:rho0}) and compare with (\ref{Eq:MFATFiniteSub}) and (\ref{eq:VarianceSubdiffusion})).

\section{Second model: resetting of the position}
\begin{figure*}[t]
\includegraphics[width=0.49\linewidth]{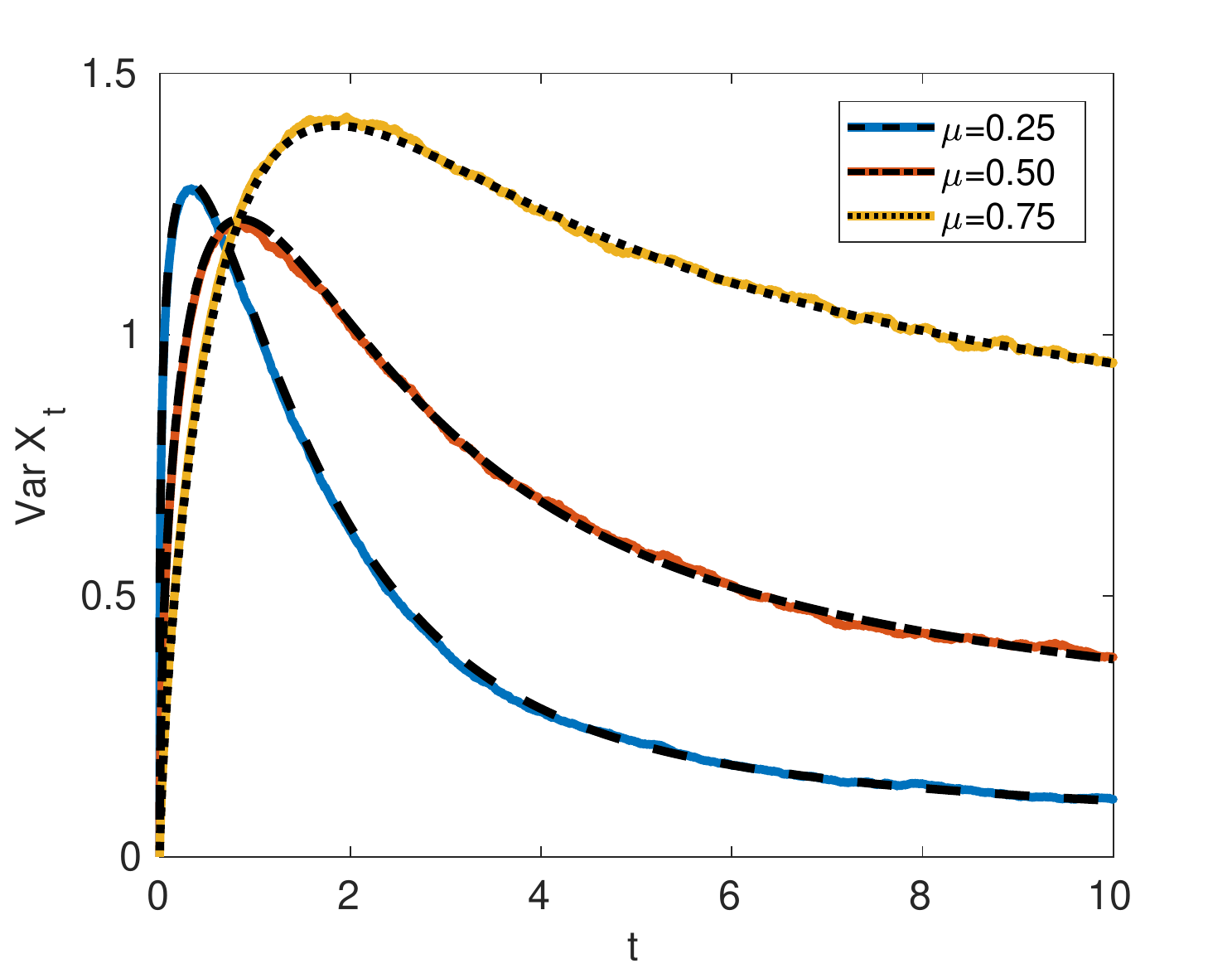}
\includegraphics[width=0.49\linewidth]{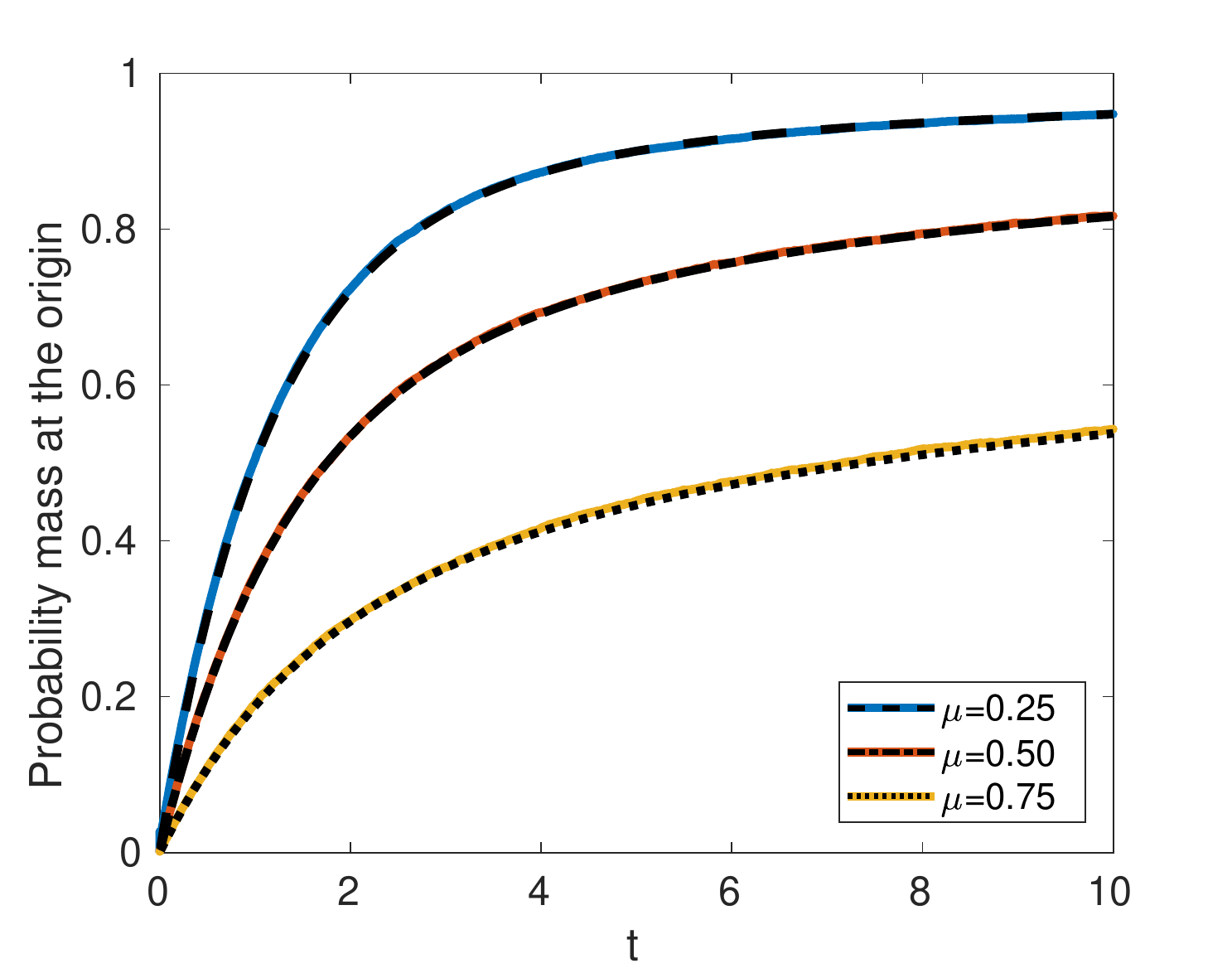}
\caption{Comparisons of variances (left) and $p_{\mathrm{origin}}$ (right) 
as a functions of time obtained with 
analytical prediction (Eqs.~(\ref{Eq:model2variance2}) and (\ref{eq:porigin}), 
dashed black lines. Calculations of $\,_1F_1$ were performed with the help 
of the code published in \cite{pearson2009computation}) 
and with stochastic simulations (colored lines). 
Details of the simulations: 
Waiting times were generated from a power-law distribution 
with $\mu=\{0.25, 0.5, 0.75\}$ and cut-off time scales 
$\tau_{0}=\{10^{-9},10^{-8},10^{-7}\}$, 
where $A_{\mu} = \tau_{0}^{\mu}$. 
Steps were generated from a normal distribution with 
$\sigma^2 = 2 D \tau_0^{\mu}$.
Trajectories were sampled every $\Delta t=10^{-3}$ 
and resets were introduced at 
each time step with probability $r\Delta t$. 
Number of sample trajectories: $10^5$. 
Other parameters: $D=r=1$.
}
\label{Fig:model2_numerics}
\end{figure*}
In our second model we assume that resets bring the particle back to $x=0$, 
but do not affect the waiting times, 
i.e. after the reset the particle still waits for the next jump as scheduled 
by the previously generated waiting time. 
We will refer to this resetting mechanism as soft resets. 
If the process is space homogeneous (invariant under translations), 
one can first generate the trajectory of the process without resets, 
and then introduce the resets by shifting the trajectory from the times of resets onward. 

Similarly to the standard CTRW scenario, we derive the formula for the evolution of the PDF of the process 
$W(x,t)$ by considering a set of renewal integral equations linking the propagator and an auxiliary function $Q(x,t)$, 
which describes the density of particles jumping from the position $x$ at time $t$. 
We assume that initially the distribution of the particles is described by $P_0(x)\equiv W(x,t = 0)$ 
and all particles have their waiting times generated at $t=0$ (no history). 
Resets are generated from the exponential distribution, 
thus the probability that resets do not happen up to time $t$ is given by $\Psi_R(t) = \exp(-r t)$. 
\begin{multline}
    W(x,t) = 
    P_0(x) \Psi(t) \Psi_R(t) + 
    \delta(x) \Psi(t) \left(1 - \Psi_R(t) \right) + 
\\
 +   \int\limits_{-\infty}^{\infty}d y \int\limits_0^t d\tau g(y) Q(x-y, t-\tau)\Psi(\tau)\Psi_R(\tau) + 
 \\
+ \delta(x) \int\limits_{-\infty}^{\infty}dz\int\limits_{-\infty}^{\infty}d y \int\limits_0^t d\tau 
    g(y) Q(x-y,t-\tau)\Psi(\tau)\left( 1 - \Psi_R(\tau)\right),
    \label{eq:WxtIntegral}
\end{multline}
where the four terms correspond to the following cases: 
1) no jumps and and no resets up to time $t$, 
2) no jumps and at least one reset up to time $t$, 
3) last jump at time $t-\tau$, no resets from $t-\tau$ up to $t$,
4) last jump at time $t-\tau$, at least one reset from $t-\tau$ up to $t$. 
Similarly, the equation for $Q(x,t)$ reads
\begin{multline}
    Q(x,t) = 
    P_0(x) \psi(t) \Psi_R(t) + 
    \delta(x) \psi(t) \left(1 - \Psi_R(t) \right) + 
\\
 +   \int\limits_{-\infty}^{\infty}d y \int\limits_0^t d\tau g(y) Q(x-y, t-\tau)\psi(\tau)\Psi_R(\tau) + 
 \\
+ \delta(x) \int\limits_{-\infty}^{\infty}dz\int\limits_{-\infty}^{\infty}d y \int\limits_0^t d\tau 
    g(y) Q(x-y,t-\tau)\psi(\tau)\left( 1 - \Psi_R(\tau)\right).
    \label{eq:QxtIntegral}
\end{multline}
Notice that the only change on the RHS is the usage of the waiting times PDF $\psi$ 
instead of the survival probability $\Psi$. 
We can rewrite (\ref{eq:WxtIntegral}) and (\ref{eq:QxtIntegral}) in the Fourier-Laplace space
\begin{multline}
    W(k,s) = 
    P_0(k)\Psi(s+r) + g(k) Q(k,s)\Psi(s+r) + 
    \\
    +\left(1 + Q_0(s)\right)\left(\Psi(s) - \Psi(s+r)\right)
    \label{eq:WksIntegral}
\end{multline}
\begin{multline}
    Q(k,s) = 
    P_0(k)\psi(s+r) + g(k) Q(k,s)\psi(s+r) + 
    \\
    +\left(1 + Q_0(s)\right)\left(\psi(s) - \psi(s+r)\right)
    \label{eq:QksIntegral}
\end{multline}
where we have introduced 
\begin{equation}
    Q_0(t) = \int\limits_{-\infty}^{\infty}dx Q(x,t),
    \label{eq:Q0def}
\end{equation}
which describes the total flow of particles at time $t$ independently from the position.
Since resets do not affect waiting times, we can write a simple renewal equation
\begin{equation}
    Q_0(t) = \psi(t) + \int_0^t d\tau \psi(\tau)Q_0(t-\tau),
\end{equation}
which immediately leads to 
\begin{equation}
    Q_0(s) = \frac{\psi(s)}{1 - \psi(s)}.
    \label{eq:Q0solution}
\end{equation}
Algebraic equations (\ref{eq:WksIntegral}), (\ref{eq:QksIntegral}), and (\ref{eq:Q0solution}) 
lead to the following general solution
\begin{equation}
    W(k,s) = \frac{P_0(k)\Psi(s+r)}{1 - g(k) \psi(s+r)} + 
    \frac{1}{s}\left(
    1 - \frac{\Psi(s+r)}{\Psi(s)}\frac{1 - g(k)\psi(s)}{1-g(k)\psi(s+r)}
    \right)
\end{equation}
As before, we assume that $\psi(s)\approx 1 - A_{\mu} s^{\mu}$ and $g(k) \approx 1 - D_0 k^2$, 
take the limits $A_{\mu}\to 0$ and $D_0 \to 0$ with $A_{\mu}/D_0 \to D$, and arrive at
\begin{equation}
    W(k,s) = 
    \frac{ P_0(k) + (1+r/s)^{\mu} - 1 }{r + s + D(r+s)^{1-\mu}k^2}
    + \frac{1}{s} - \frac{1}{s^{\mu}(r+s)^{1-\mu}},
    \label{eq:WksSolution}
\end{equation}
which is our central result in this section. 

As we will now show, the process posses rather peculiar qualities. 
The stationary probability is given simply by
\begin{equation}
f_s(x) = \lim_{t\to \infty}W(x,t)=\lim_{s\to 0} s W(x,s) = \delta(x),
\label{Eq:stationary2}
\end{equation}
and is the same as the initial distribution if $x_0=0$ (i.e. ${P_0(k) = 1}$). 
It does not, however, imply that the process is trivial ($X_t=0$). 
In order to show this we calculate the variance as a function of time.
As in the previous example we assume that $x_0=0$ and obtain the following 
expression for the MSD in the Laplace space
\begin{equation}
\langle X^2(s) \rangle = \frac{2 D}{s^{\mu} (s+r)}.
\label{eq:model2varianceLaplace}
\end{equation}
Inverting the Laplace transform leads to the integral expression
\begin{equation}
\langle X^2(t) \rangle = 
\frac{2 D}{\Gamma(\mu)}e^{-r t}\int\limits_0^t d\tau \tau^{\mu-1}e^{r \tau},
\label{Eq:model2variance}
\end{equation}
which can be represented in terms of the confluent hypergeometric function of the 
first kind
\begin{equation}
\langle X^2(t) \rangle = 
\frac{2 D}{\Gamma(\mu + 1)}t^{\mu} e^{-r t} \,_1F_1(\mu,\mu+1,r t).
\label{Eq:model2variance2}
\end{equation}
The MSD as a function of time admits maximum value at $t^*=z_0/r$, 
where $z_0$ is the positive solution of the equation
\begin{equation}
\int\limits_0^{z_0} z^{\mu-1} e^{z} dz = 
z_0^{\mu-1} e^{z_0},
\end{equation}
which can rewritten in terms of a zero of the confluent hypergeometric function
\begin{equation}
{}_1F_1(\mu-1;\mu;z_0)=0.
\end{equation}
The maximum value of the MSD takes the form
\begin{equation}
\max_{t} \langle X^2(t) \rangle 
= \frac{2 D}{r^{\mu}\Gamma(\mu)} z_0(\mu)^{\mu-1}.
\end{equation}
By expanding (\ref{eq:model2varianceLaplace}) we see that 
$\langle X^2(s)\rangle \propto s^{-\mu}$ for $s\ll r$, 
which translates to $\langle X^2(t)\rangle \propto 1/t^{1-\mu}$ 
for large $t$. 
The stationary state is approached with a slow, algebraic decay, 
which gets slower with larger values of $\mu$, 
see Fig.~\ref{Fig:model2_numerics}. 

\begin{figure}[t]
\includegraphics[width=1.0\linewidth]{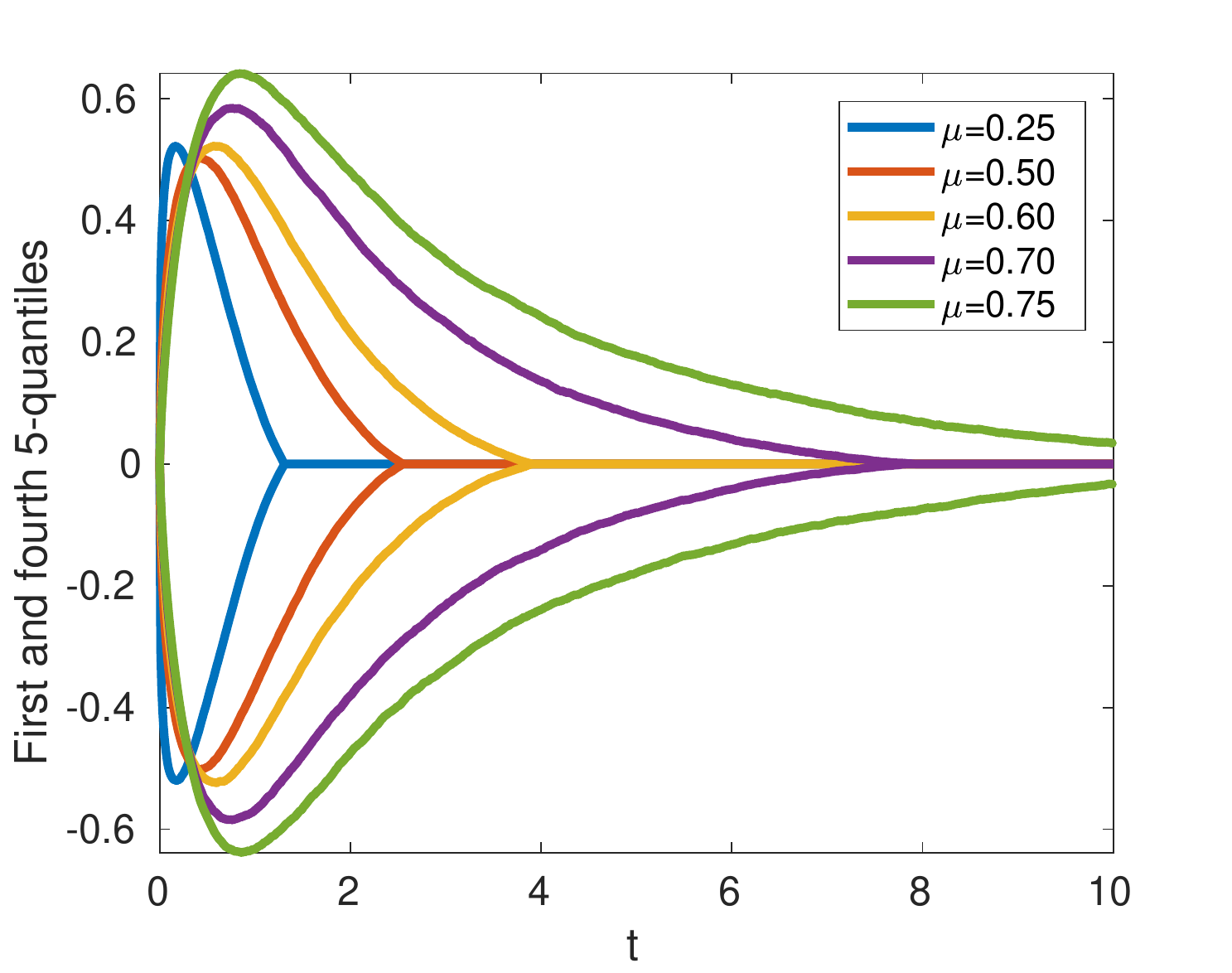}
\caption{
Quantile lines of L\'evy flights with power-law waiting 
times corresponding to (\ref{eq:WksLFSolution}) with $\alpha=1$ 
\lk{and different values of $\mu$ (increasing from left to right)}. 
Results were obtained by means of stochastic simulations. 
Steps were generated from a Cauchy distribution. 
Trajectories were sampled every $\Delta t=0.01$ 
and resets were introduced at 
each time step with probability $r\Delta t$. 
Number of sample trajectories: $10^6$. 
Other parameters: $D=r=1$. 
}
\label{Fig:model2cauchy}
\end{figure}

To sum up, the initial and the stationary distribution are the same, 
with a nontrivial transient behavior. 
This bears a resemblance to homoclinic orbits in dynamical systems 
\cite{strogatz2014nonlinear} and 
is related to the stationary state splitting observed recently in 
a similar model \cite{shkilev2017continuous}. 
One may think that this non-monotonic behavior is induced by subdiffusion. 
In order to show that this is not the case let us modify the model 
by introducing a power-law, heavy-tailed distribution of jumps, 
i.e. $g(k) \approx 1 - D_0 |k|^{\alpha}$, with $\alpha\in(0,2)$.
The corresponding process without resets is described by the following FFPE
\begin{equation}
    \partial_t W_0(x,t|x_0) 
=
\,_0D{_t^{1-\mu}} D \partial_{|x|}^{\alpha} W_0(x,t|x_0),
\label{eq:W0LFequation}
\end{equation}
where $\partial_{|x|}^{\alpha}$ denotes the Riesz fractional space operator \cite{metzler2000random}. 
Sample paths of the process are discontinuous, 
and its increments are described by L\'evy $\alpha$-stable distribution \cite{fellerBook}, 
hence the name L\'evy flights \cite{mandelbrotBook,hughes1981random}. 
Additionally, the variance of the position is infinite for any $\alpha<2$. 
Nevertheless, one can still differentiate between superdiffusive and subdiffusive cases using quantiles 
(we could avoid this difficulty by using L\'evy walks, 
but we choose L\'evy flights for the sake of simplicity of the calculations). 
In the case of the process with soft resets, the same calculations as before lead to the following formula 
\begin{equation}
    W(k,s) = 
    \frac{ P_0(k) + (1+r/s)^{\mu} - 1 }{r + s + D(r+s)^{1-\mu}|k|^{\alpha}}
    + \frac{1}{s} - \frac{1}{s^{\mu}(r+s)^{1-\mu}},
    \label{eq:WksLFSolution}
\end{equation}
The underlying process without resets is self-similar 
and its quantiles scale with time as $\mu/\alpha$. 
Therefore the process behaves subdiffusively for $2 \mu<\alpha$ 
and superdiffusively for $2\mu>\alpha$. 
However, it is easy to show that the non-monotonic behavior of the 
process with position resets appears independently of $\alpha$, 
since the stationary probability
\begin{equation}
f_s(x) = \lim_{t\to \infty}W(x,t)=\delta(x),
\end{equation}
which holds for $\mu<1$ and any $\alpha$. 
This time we cannot use the MSD as a proxy to see the dispersion in time, 
but we can plot quantile lines to confirm the non-monotonic behavior, 
see Fig.~\ref{Fig:model2cauchy}. 
Thus, it is evident that the non-monotonic behavior of the process 
relies on the trapping with power-law waiting times.  
\lk{For the same reason no such non-monotonic behavior 
is observed in the case of subdiffusion with hard resets, 
wherein power-law waiting times are effectively exponentially truncated. 
}
We hypothesize that this is related to the weak ergodicity breaking, 
i.e. we expect the non-monotonic behavior will not appear in 
the fractional Brownian motion or random walks on fractals. 
If this is the case, one could use the resetting as a computational method to
assess whether the process at hand is weakly non-ergodic. 

Fig.~\ref{Fig:model2cauchy} illustrates another 
interesting feature of the CTRW with soft resets: 
Quantile lines converge to zero in a finite time, which 
means that even for $t<\infty$ there is a finite probability $p_{\mathrm{origin}}(t)$
that the particle is exactly at the origin. 
This probability corresponds to the singular part of $W(x,t)$ and 
can be calculated as $\lim_{k\to\infty} W(k,s)$, 
which for any $g(x)$ that is continuous at $x=0$ reads
\begin{equation}
    p_{\mathrm{origin}}(s) = \frac{1}{s} - \frac{1}{s^{\mu}(r+s)^{1-\mu}},
\end{equation}
which can be represented in terms of the confluent hypergeometric function
\begin{equation}
    p_{\mathrm{origin}}(t) = 1 - \,_1F_1(1-\mu;1;-r t) = 1 - e^{-r t} \,_1F_1(\mu;1;r t). 
    \label{eq:porigin}
\end{equation}
Fig.~\ref{Fig:model2_numerics} shows the behavior of 
$p_{\mathrm{origin}}(t)$ for different values of $\mu$. 
Our analytical predictions (\ref{eq:porigin}) are fully 
corroborated by the stochastic simulations. 
Note that for any value of $\mu<1$ the density is concentrated in the atom in the 
limit of $t{\to\infty}$. 
However, the closer the value of $\mu$ to $1$, the slower the convergence, 
in line with the asymptotic ($t\gg 1/r$) formula 
\begin{equation}
    1 - p_{\mathrm{origin}}(t) \propto 1/t^{1-\mu}.
\end{equation}

We now turn our attention to the FAT statistics in the model of subdiffusion with soft resets. 
Since resets do not affect waiting times for random jumps and 
for $\mu<1$ these waiting times have infinite expected value,
we expect the MFAT to be infinite for $\mu<1$. 
Moreover, since at any given moment some particles are trapped at the origin 
and the proportion of trapped particles increases with time, 
one may expect that there is a finite probability that the target will never be found. 
This is, surprisingly, not the case. 
By means of the general formula for the first passage times density (\ref{eq:WandPalg}) 
together with (\ref{eq:WksSolution}) we find
\begin{equation}
    \rho(s) = \frac{1}{1 + \left(\frac{s}{s+r}\right)^{\mu}\left(\exp\left((r+s)^{\mu/2}|x_0|/\sqrt{D}\right) - 1 )\right)},
\end{equation}
which can be rewritten as
\begin{equation}
    \rho(s) = \frac{1}{\left(\frac{s}{s+r}\right)^{\mu}(\rho_0(s+r))^{-1} + 1 -  \left(\frac{s}{s+r}\right)^{\mu}  },
    \label{eq:pfa2}
\end{equation}
compare to (\ref{eq:pfa1}).
In contrast to hard resets, soft resets do not lead to finite MFAT--as expected 
$\langle T_r\rangle = \infty$ for $\mu<1$. 
However, $\lim_{t\to \infty}S(t)=0$, i.e. the particle reaches the target almost surely. 
The tail of $\rho(t)$ behaves like $1/t^{1+\mu}$ and is lighter than the tail of 
the process without resets $1/t^{1+\mu/2}$. 

\section{External potentials: drift-subdiffusion with resets}
We can introduce external potentials into the 
subdiffusive CTRW and the corresponding FFPE in different ways: 
trapping (waiting times) may directly affect both random and potential-dependent terms \cite{metzler1999anomalous,metzler2000random}. 
The FFPE takes the form 
\begin{equation}
    \partial_t W_0(x,t|x_0) 
    =
    {}_0 D{_t^{1-\mu}}\left( 
    D \partial_{xx}^2 W_0(x,t|x_0)
    +
    \partial_x \left( U'(x) W_0(x,t|x_0)\right)
    \right),
    \label{eq:W0-FFPE-drift1}
\end{equation}
where $U(x)$ is the time-independent external potential and 
${F(x)\propto-U'(x)}$ is the corresponding external force 
(for a time-dependent generalization see \cite{henry2010fractional}). 
In an alternative approach, waiting times may be affected by the external force, 
leading to another, more complicated form of the FFPE, 
in which both terms depend on the external force \cite{fedotov2015subdiffusion}.

Here we explore (\ref{eq:W0-FFPE-drift1}) 
with a constant force $f = -U'(x)$, 
i.e. $g(k)\approx 1 + i k f_0 - D_0 k^2$, 
where $f_0\to0$, $D_0\to0$, $f_0/A_{\mu}\to f$, and $D_0/A_{\mu}\to D$.
The corresponding free propagator 
(without absorbing boundaries) takes the form
\begin{equation}
    W_0(k,s|x_0) = 
    \frac{e^{i k x_0}}{s + D s^{1-\mu} k^2 - i k f s^{1-\mu} }.
    \label{eq:W0drift}
\end{equation}
In order to calculate the FAT statistics we again place an absorbing 
target at $x=0$ and apply the method of images 
with a linear combination of solutions (\ref{eq:W0drift}) of (\ref{eq:W0-FFPE-drift1}) 
starting from $\pm x_0$. 
For simplicity of the notation and without loss of generality we assume that the process starts from $x_0>0$.
The formula
\begin{equation}
    G_0(x,s) = \frac{c^2 e^{\bar{f}(x-x_0)}
    }{2 s \sqrt{c^2 + \bar{f}^2}}
    \left(
    e^{
    -\sqrt{\bar{f}^2 + c^2}|x-x_0|
    }
    -
    e^{
    -\sqrt{\bar{f}^2+c^2}|x+x_0|
    }
    \right),
\end{equation}
with $c=\sqrt{s^{\mu}/D}$, 
is a straightforward generalization the known case of diffusion \cite{fellerBook}. 
The corresponding survival probability reads
\begin{equation}
    S_0(s) = \frac{1}{s}\left(
    1 - \exp{\left(-\left(\bar{f} + \sqrt{\bar{f}^2 + c^2}\right)x_0\right)}
    \right).
\end{equation}
The infinite-time survival depends on the sign of $f$ and is equal to $0$ 
if $f<0$ (target is found with probability $1$), whereas for positive force 
the target is found with probability $\exp(-2 \bar{f} x_0)$. 
The MFAT is finite if and only if $f<0$ and $\mu=1$ and in this case is 
given by the ballistic time $\langle T_r \rangle = x_0/|f|$.

The propagator of the corresponding model with hard resets 
to the origin can be calculated from (\ref{eq:WfromW0}), 
and the corresponding stationary distribution reads
\begin{equation}
   f_s(x) = 
   \frac{1}{2} 
   \frac{a^2}{\sqrt{a^2 + \bar{f}^2}} 
   \exp\left(\bar{f}x - \sqrt{\bar{f}^2 + a^2} |x|\right),
\end{equation}
where $\bar{f}=\frac{f}{2D}$ and $a = \sqrt{r^{\mu}/D}$.
The corresponding MFAT in the process with hard resets to the initial position $x_0$ 
can be calculated with (\ref{eq:exponentialMeanT}) and reads
\begin{equation}
    \langle T_r \rangle = 
    \frac{1}{r}\left(
        \exp{\left(
            \left(\bar{f} + \sqrt{\bar{f}^2 + a^2}\right)
            x_0\right)
            }
            -1
    \right)
\end{equation}
It has recently been shown that whether or not resets can lower the 
MFAT of diffusion with drift is controlled by the P\'eclet number \cite{ray2018p}. 
Our results show that hard resets can always help for $\mu<1$, 
since in this case the process without resetting has infinite MFAT.

The propagator of the model with soft resets takes the form
\begin{multline}
W(k,s|x_0) = 
\frac{ \exp(i k x_0) + (1+r/s)^{\mu} - 1 }{r + s + D(r+s)^{1-\mu}k^2 - i f (r+s)^{1-\mu}k} +
    \\
    + \frac{1}{s} - \frac{1}{s^{\mu}(r+s)^{1-\mu}}
.
\label{Eq:Propagator2drift}
\end{multline}
As before, Model II exhibits trivial stationary distribution 
${f_s(x) = \delta(x)}$ with nontrivial transient. 
In this case the non-monotonic behavior can be already observed in the first moment
\begin{equation}
    \langle X(s) \rangle = 
    \frac{x_0}{r+s} + \frac{f}{s^{\mu}(r+s)},
\end{equation}
where the second term behaves like the variance in the model 
without the drift term (\ref{eq:model2varianceLaplace}). 
One can easily verify that in the case of soft resets drift does not change the
main conclusions regarding the FAT statistics: 
yet again the particle almost surely reaches the target, 
but the MFAT is infinite.

\section{Conclusions}
Two models of subdiffusion with stochastic resetting have been presented. 
Their statistical properties, including stationary distributions, 
transient behavior of the probability distribution functions, 
as well as first two moments of first arrival time have been calculated. 
{While the subdiffusion with hard resets provide a rather 
straightforward generalization of diffusion with stochastic resetting, 
soft resets lead to quite peculiar features of the process, 
including non-monotonic behavior of the MSD and non-intuitive properties 
of the first passage time statistics.
}
In the following we list a few possible applications of these results. 

A number of models of molecular motors dynamics based on ratchet mechanism 
in a thermally fluctuating environment have been proposed before
\cite{vale1990protein,cordova1992dynamics,astumian1994fluctuation,reimann2002brownian,astumian2002brownian,Klumpp2005,RevModPhys.81.387,lisowski2015cargo}. 
Since the cytoplasm in the living cells is crowded, 
the transport inside of them is observed to be subdiffusive 
(cf. Section \ref{sec:introduction}). Therefore, some authors have devised 
molecular motors models in subdiffusive environments 
\cite{PhysRevLett.85.5655,Zaid2009710,PhysRevE.80.011912,goychuk2014molecular,goychuk2015anomalous}. 
Our first model of subdiffusion with stochastic resetting may provide 
an useful building block in such models, wherein resetting would describe 
binding and unbinding of molecular motors from microtubules 
or disappearance and appearance of particles due 
to chemical reactions \cite{rotbart2015michaelis,Reuveni2016}.

The second model introduced in this work may be used as a starting point 
to construct novel resampling statistical methods 
\cite{efron1982jackknife,westfall1993resampling,sauerbrei1999use,yekutieli1999resampling,mackinnon2004confidence,howell2012statistical,good2013permutation}, 
e.g. for an estimation of the tail index or testing hypothesis whether 
the observed trajectories are subdiffusive \cite{meroz2015toolbox}. 
The idea is to create auxiliary trajectories by randomly resetting positions 
in the original data, with the hope of the new data being easier to handle, 
especially when a limited number of trajectories is given. 
The details of the algorithm and whether this hope is justified 
will be the subject of further research. 

\begin{acknowledgments}
This work was supported by RIKEN Center for Brain Science. 
\end{acknowledgments}
\bibliography{citations}
\bibliographystyle{phjcp}
\end{document}